\documentclass[aps,preprint]{revtex4}
\usepackage{epsfig, amssymb}

\begin{document}


\title{{\large {\bf  The Commercial Stainless Steel Tube Enveloping Technique for ${\bf MgB_2}$}
}}

\author{H. B. Lee, B. J. Kim, Y. C. Kim$^{\rm a)}$
\footnotetext{a) Corresponding author: Tel 82 - 51 - 510 - 2224, e-mail : yckim@pusan.ac.kr }} 
\address{\ Department of Physics, Pusan National University, Busan 609-735, Korea}
\author{D. Y. Jeong}
\address{\ Korea Electrotechnology Research Institute, Changwon 641-120, Korea}

\begin{abstract} 
  A commercial stainless steel tube was employed to fabricate {\rm MgB$_2$}. 
  The specimen  was prepared by a stoichiometric mixture of Mg and B inside the stainless 
  steel tube. The specimen was sintered for 2 hours at 920$^{o}$C. 
  X-ray spectra showed there were no second phases like MgO. The transition temperature of the 
  specimen was 37.5 K with a sharp transition width within 1K.  The specimen showed a good 
  connection between grains and critical current density as calculated with the Bean model is more 
than
  бн10$^{5}$ A/cm$^{2}$ in the 20 K and zero field. \\ 

PACS numbers : 74.60.-w; 74.70.Ad; 89.20.Bb

\end{abstract}


\maketitle

\vspace{0.5cm}

\newpage

   After Akimisu announced the discovery of 39K-temperature superconductivity in {\rm MgB$_2$}, it 
has 
 stimulated considerable interests in superconducting research 
groups.\cite{Akimisu}\space\space\space  
 In the early study, it was reported  that {\rm MgB$_2$} superconducting phase was fabricated 
 under the condition of a high temperature and  high pressure.\cite{Monteverde}\space\space\space 
But
 it was soon revealed that high pressure was needed to prevent Mg from escaping 
 from the base materials.\cite{Kang,Eom}\space\space\space It is because Mg is a too 
 volatile material. Many groups reported that if Mg did not escape from the base materials
 in the elevated temperature, {\rm MgB$_2$} was fabricated.\cite{Jin,Canfield}\space\space\space 

   An effective  way  to prevent the escape of some kinds of  elements in the base
 materials during heat treatment is envelope treating. It has been mainly used when 
low-melting-point and
 high-evaporating materials are heat treated, especially Tl, Hg, Mg, 
etc.\cite{Eom,Foong,Yan}\space\space\space 
 From now on,  refractory metals and quartz tube are mainly used for materials of 
 enveloping.\cite{Kang,Eom}\space\space\space Ta and Nb are candidate of refractory metals. These 
materials have a high melting 
 point and a strong Mg corrosion resistance.\cite{ASM Handbook Committee}\space\space\space 
 But in  the elevated temperature, these materials are like to oxidize.  Because of the high 
reactivity of Mg with oxygen, even a small 
 hole can make it very bad to fabricate {\rm MgB$_2$}. To prevent the surface from oxidizing, we 
have to treat 
 specimens in inert gas or in a vacuum state. This is another problem of rising cost, 
 especially fast-heat and fast-cool treating.
 Recently many groups used to envelope a quartz tube to outside Ta envelope once more
 and evacuate quartz tube to avoid surface oxidation of Ta.\cite{Kang,Canfield}\space\space\space

  The quartz tube for the envelope treating was used for a rather lower fabricating temperature than 
for {\rm MgB$_2$},
  like Hg-based and Tl-based superconductors.\cite{Eom,Foong,Yan}\space\space\space Use of the 
quartz tube for fabricating the
 {\rm MgB$_2$} superconductor was restrained owing to softening of the quartz tube in the elevated
 temperature(around 900${\rm ^oC}$) and the reaction of Si with Mg. So the quartz tube cannot be 
used but rather 
 a second material to avoid surface oxidation in fabricating the {\rm MgB$_2$} superconcuctor. 
 
   In the case of short films with the {\rm MgB$_2$} superconductor, the refractory metals and the 
quartz tube 
 for the envelope are believable, but rather inconvenient for Mg's affinity for a oxygen and 
 high cost. With respect to fabricating long-length wires and films with the {\rm MgB$_2$} 
superconductor, 
 these are more difficult to fabricate owing to the possibility of breakage in using the quartz tube 
 and the cost of fabricatation.

  So, there are deep demands that are an economical and easy treatment method for envelope 
technology. 
 Now, we have  developed  the Commercial Stainless Steel Tube Envelope Technique(COSSET) for the 
{\rm MgB$_2$} superconductor. The COSSET is easy, 
 economical and believable. The COSSET is hard enough to be in a longer heat treatment of the 
elevated
 temperature and is not 
 breakable like the quartz tube and does not need special environments like Ta and Nb. It will be 
 very useful  for long length wires and films of {\rm MgB$_2$}. 
  
   The starting materials are Mg(99.9\% powder)and B(96.6\% amorphous powder). 
 The sample of {\rm MgB$_2$} was prepared in several steps. Mixed Mg and B stoichiometry was
 finely ground, then pressed into a pellet 10mm in diameter. Also, an 8m-long stainless steel(304) 
tube was cut into a 10cm piece. One side of the 10cm 
 long tube forged and welded and Fe plate was inserted into the stainless steel tube. The pellet was 
put
 on the Fe plate. The pellet had been heat treated  at 300${\rm ^oC}$ for 1hr to harden it 
 before insertion into the stainless steel tube. Excess Mg was put under the Fe plate and  the other 
side of the 
 stainless steel tube was forged and put into a high-purity Ar gas in the stainless steel tube, and 
which was welded. 
 Finally, it was heat treated at 920${\rm ^oC}$ for 2hrs using the fast-heat and fast-cool method in 
air.\cite{Yan}\space\space\space  

  Figure 1 shows the XRD pattern of the {\rm MgB$_2$} bulk sample heat treated at 920${\rm ^oC}$ for 
2hrs. 
 There are no second phases like with MgO in the XRD pattern. Second phases like MgO itself 
 do not harm the superconductivity, because  it acts as a pinning center. 
 But during the process to fabricate {\rm MgB$_2$}, the existence of MgO means that base 
materials(Mg and B) have  
 reacted with outside oxygen. That would make the superconducting parts small and the 
 non-superconducting parts large, and would drop the confidence of superconductivity. 
 By comparing the COSSET with other processes such as  high 
pressure,\cite{Takano,Jung}\space\space\space
 PIT(powder in tube)\cite{Jin}\space\space\space and simple wrapping 
 in iron plate,\cite{Dou}\space\space\space  the COSSET efficiently restrained the oxygen from 
 their source (the outside of the envelope). This would lead us to believe that the sample has good 
property of {\rm MgB$_2$} in the COSSET. 
  
   Figure 2 shows the SEM photographs of the {\rm MgB$_2$} bulk sample. Part(a) shows the surface of 
the pellet,
 and part(b) shows the inside of the pellet.  Part(b) shows different results from {\rm MgB$_2$} 
fabricated by other processes such as high pressure, PIT or simple wrapping. 
 The inside of the pellet  was condensed by the COSSET during the fabrication of {\rm MgB$_2$}. It 
is not 
 surprising that {\rm MgB$_2$} was condensed, because each Mg and B were composed of one {\rm 
MgB$_2$}.
 The condensing of the specimen would give rise to a porosity. We have not found a micro structure 
 like this in other reports. And we could also learn that the {\rm MgB$_2$} grains
 are connected to each other by some kinds of chains. This is another special aspect of the COSSET.

  Generally when one fabricates {\rm MgB$_2$} by another process, an excess Mg is needed. 
 After fabricating {\rm MgB$_2$}, the excess Mg can be harmful if not finely dispersed.
 Unlike other processes, in the COSSET excess Mg was added separately to the {\rm MgB$_2$} pellet 
 and there was no non-dispersed Mg in the {\rm MgB$_2$} superconducting phase. In other words, 
 because the excess Mg is supposed to take part in the reaction by a vapor through the porosity, 
 the non-dispersed excess Mg is almost non-existent in the {\rm MgB$_2$} superconducting phase.
   
    The resistance versus temperature curve and the magnetic susceptibility versus temperature curve 
of the 
 {\rm MgB$_2$} bulk sample are shown in Fig. 3. It is clear that the superconducting
 transition temperature in the resistance vs temperature curve at 50mA was about 
 37.5K and the transition was sharp and the transition width was within 1K, 
 indicating the good quality of the sample.
  The magnetic susceptibility for the sample was measured at 5 Oe.
 The magnetic susceptibility date shows that the superconducting transition width was 
 also within 1K.  The decreased field cooling signal in the magnetic susceptibility 
 for the sample indicates that the flux pinning was greatly enhanced and suggests 
 a higher possibility of high current superconducting applications in the bulk form 
 and films by the COSSET. The transition temperature of the sample was also about  37.5K. 
  
   The M-H curves in Fig. 4 were measured in the temperature region from 5K to 35K.  
 The symmetry in the increasing and the decreasing field branches was good. This means
 that the contribution of the bulk pinning dominated. On the other hand, the 
 shapes of the M-H curve in the Fig. 4 have a good similarity to each other. 
 One remarkable feature is that flux jumping was shown up to the temperature of 15K. 
 Flux jumping for {\rm MgB$_2$}, which refers to a sudden dissipative 
 rearrangement of magnetic flux within a superconductor, was reported by many groups, especially 
bulk samples.\cite{Jin,Dou,Zhao}\space\space\space  The flux jumps of the sample 
 by the COSSET are rather different from ones of other processes. Most reports show that flux jumps 
appear within the temperature of 10K.
  It was well known that impurities and second phases existent in the
 specimen would influence flux jumps. The impurities and second phases pin the flux 
 and stop the flux jumps. This is also the reason that the films with the {\rm MgB$_2$} 
superconductor are little affected 
 by flux jumps. In our experiments, there is no proof of second phases, and it is considered 
 that this fact increases the temperature of flux jumps.
  
   The critical current density(Jc) versus applied fields curves of the {\rm MgB$_2$} sample were 
obtained from the Bean  model(the formula Jc=15$\Delta$M/a(1-a/3b), where 2a is the sample 
thickness and 2b is the sample width ) and are shown in the Fig. 5.
 Jc reaches more than 1$\times$10$^{5}$A/cm$^{2}$ at the temperature of 20K in zero field and 
3.3$\times$10$^{4}$A/cm$^{2}$ at 20K in 1T.
 These values of the sample are as much as or more than ones of other methods in spite of the 
COSSET's benefits and the high porosity of the sample.
   
   In summary, by using
the
Commercial
Stainless
Steel
Envelope
Technique
(COSSET),
we could
successfully
synthesize
high-quality
{\rm
MgB$_2$}
for which
the
critical
temperature
was
37.5K.
The {\rm
MgB$_2$}
has
high-porosity
structure
and there
is no
proof of
the MgO
phase.
The
magnetic
transport
J$_c$ of
the
sample is
as much
as or
more than
that of
other
methods
in spite
of the
COSSET's
benefits
and the
high
porosity
of the
sample.
The
COSSET
will be a
good
method
for the
practical
application
of {\rm
MgB$_2$}.

\begin{acknowledgments} 
 
  This work was supported by grant No. R14-2002-029-01000-0
from ABRL Program of the Korea Science and Engineering Foundation.

\end{acknowledgments}


\bigskip

\newpage \centerline{\bf FIGURE CAPTIONS}
\renewcommand{\theenumi}{FIG.~1}
\begin{enumerate}
\item {The XRD pattern of the MgB$_{2}$ bulk sample by the COSSET. There is no MgO peak in this 
pattern.}
\end{enumerate}

\renewcommand{\theenumi}{FIG.~2}
\begin{enumerate}
\item  {The scanning electron micrograph of the MgB$_{2}$ bulk sample by the COSSET: 
(a) on surface and (b) inside.}
\end{enumerate}

\renewcommand{\theenumi}{FIG.~3}
\begin{enumerate}
\item  {The magnetic susceptibility versus temperature curve of the MgB$_{2}$ bulk sample by the 
COSSET. The inset shows  resistance versus temperature dependance for the same sample.} 
\end{enumerate}

\renewcommand{\theenumi}{FIG.~4}
\begin{enumerate}
\item  {The M-H curves for the MgB$_{2}$ bulk sample  by the COSSET.}
\end{enumerate}

\renewcommand{\theenumi}{FIG.~5}
\begin{enumerate}
\item  {The critical current density versus magnetic field curves of 5 K ${\rm <}$ T ${\rm <}$ 35 K 
for the MgB$_{2}$ bulk sample  by the COSSET.}
\end{enumerate}

\newpage


\begin{figure}
\vspace{5cm}
\begin{center}
\epsfig{figure=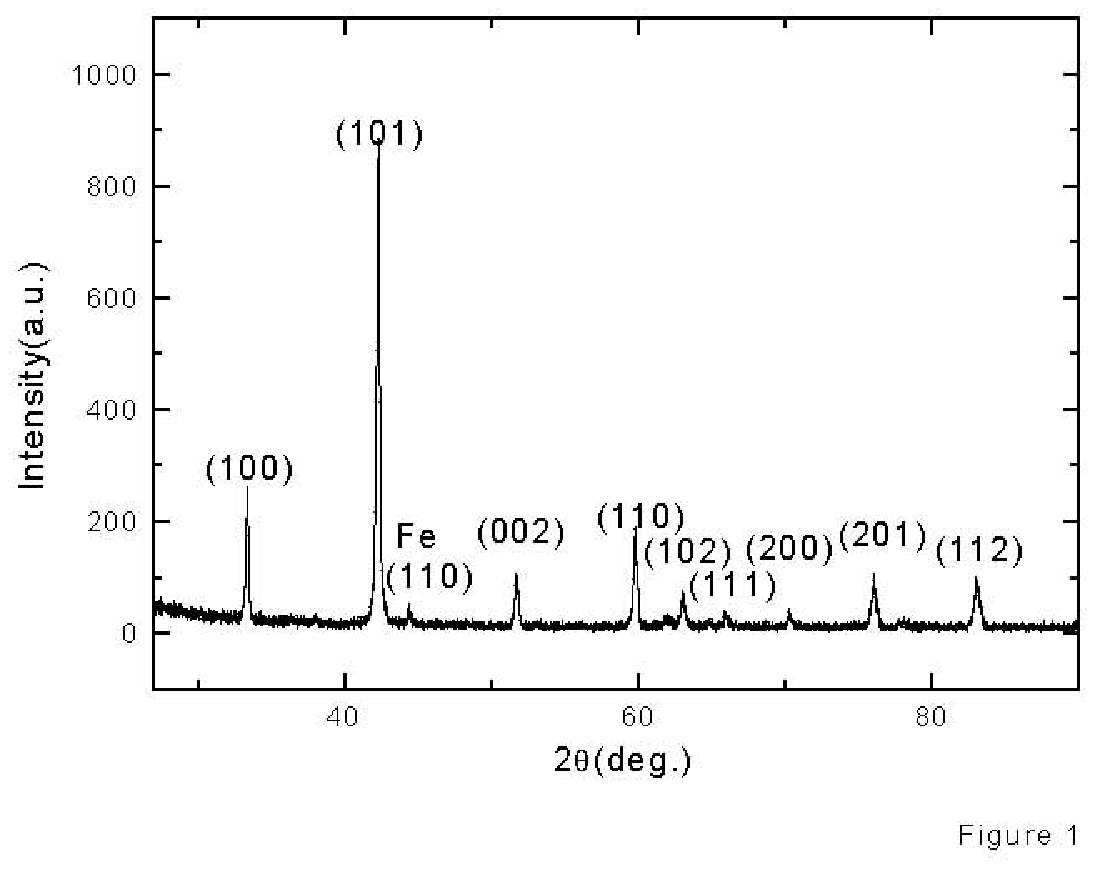,width=15cm}
\end{center}
\caption{\label{Fig11} The XRD pattern of the MgB$_{2}$ bulk sample by the COSSET. There is no MgO 
peak in this pattern.}
\end{figure}

\newpage

\begin{figure}
\vspace{3cm}
\begin{center}
\epsfig{figure=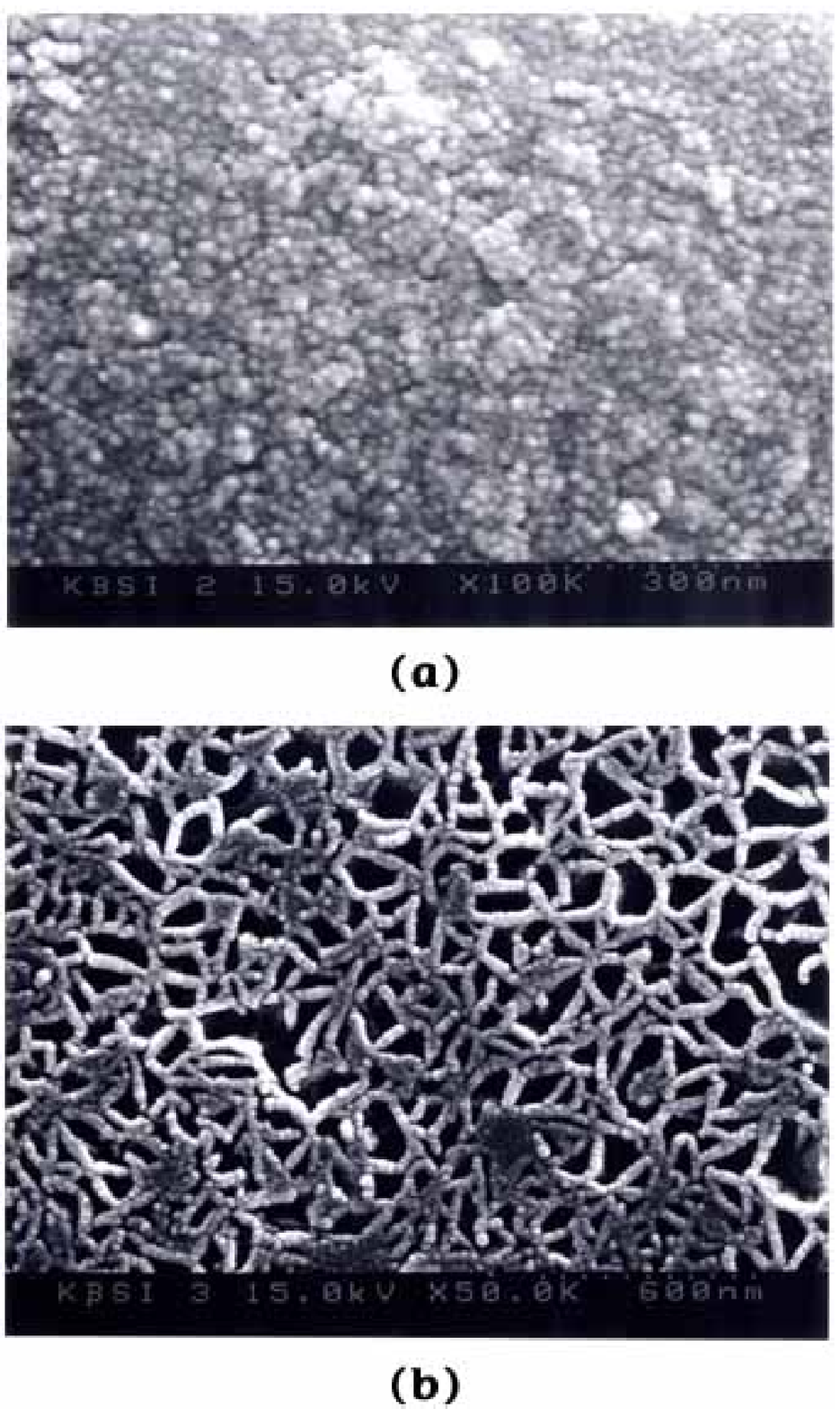,width=10cm}
\end{center}
\caption{\label{Fig12}The scanning electron micrograph of the MgB$_{2}$ bulk sample by the COSSET:
(a) on surface and (b) inside.}
\end{figure}

\newpage

\begin{figure}
\vspace{5cm}
\begin{center}
\epsfig{figure=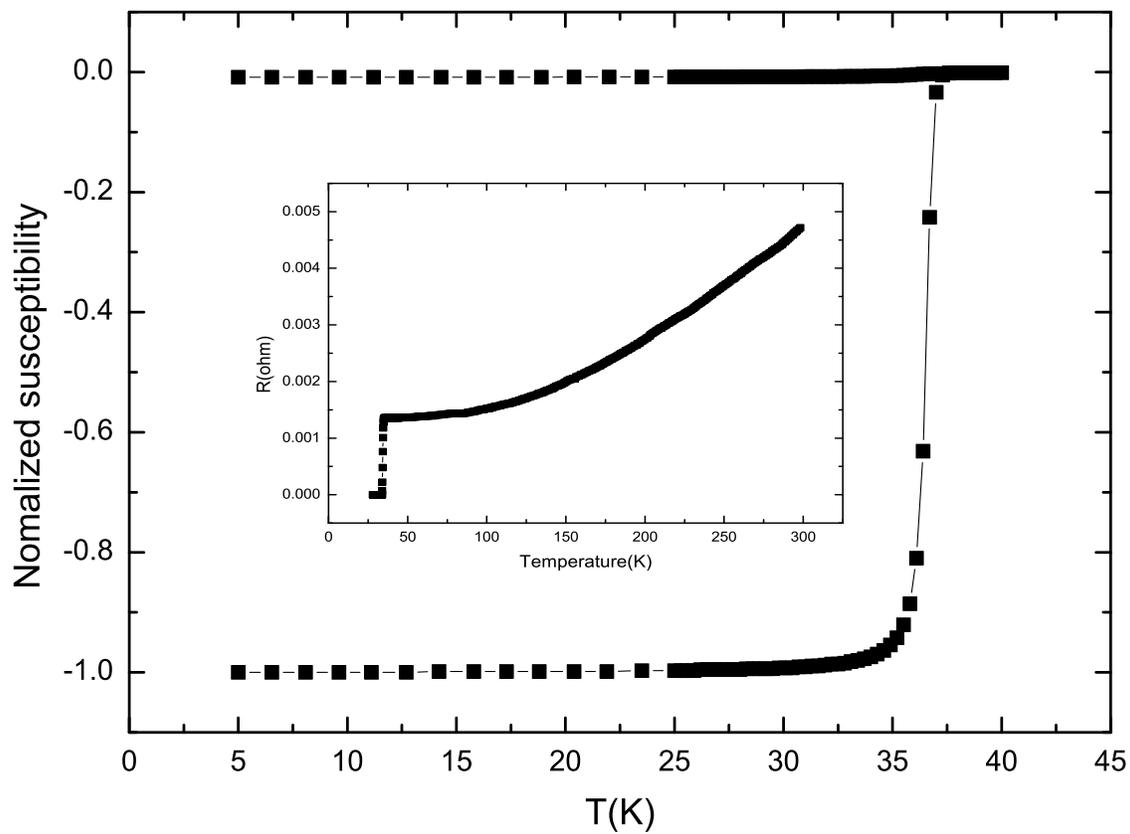,width=15cm}
\end{center}
\caption{\label{Fig13}The magnetic susceptibility versus temperature curve of the MgB$_{2}$ bulk 
sample by the COSSET. The inset shows  resistance versus temperature dependance for the same 
sample.}
\end{figure}

\newpage

\begin{figure}
\vspace{5cm}
\begin{center}
\epsfig{figure=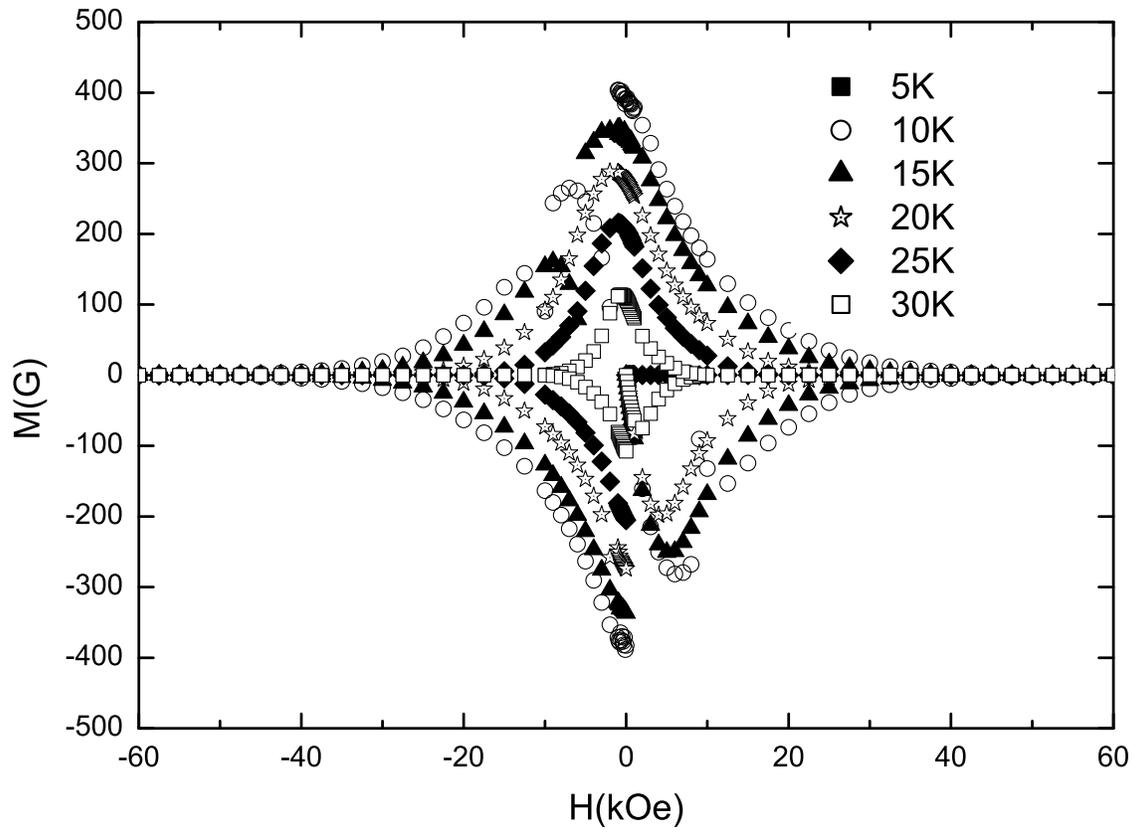,width=15cm}
\end{center}
\caption{\label{Fig14}The M-H curves for the MgB$_{2}$ bulk sample  by the COSSET.}
\end{figure}

\newpage

\begin{figure}
\vspace{5cm}
\begin{center}
\epsfig{figure=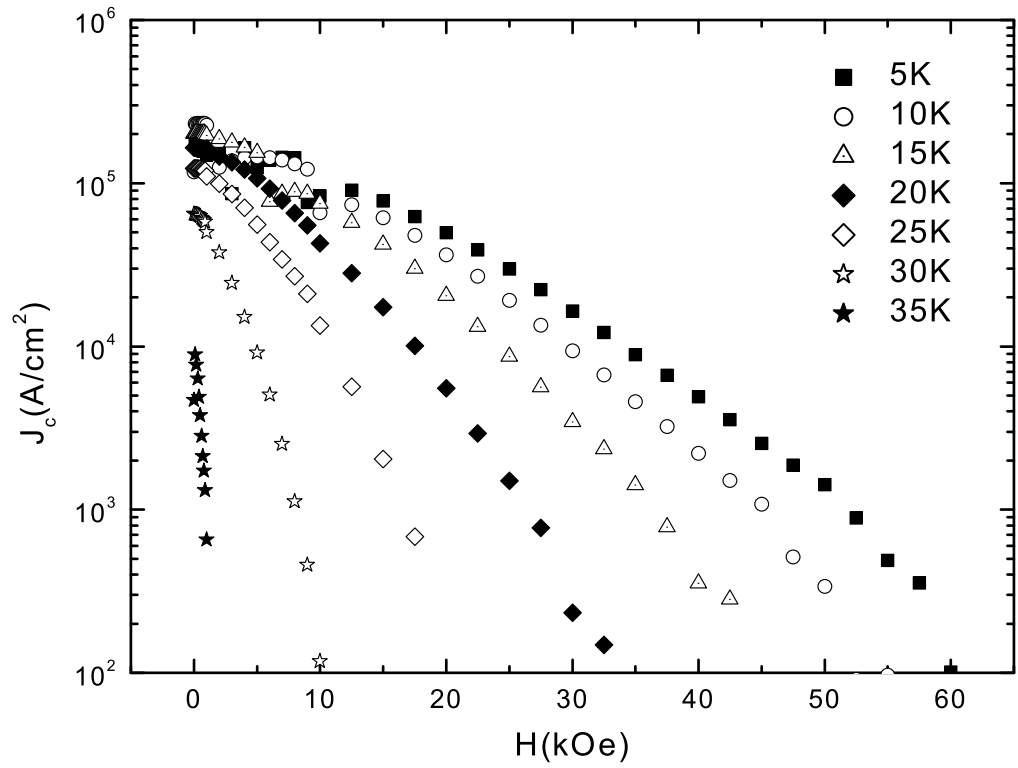,width=15cm}
\end{center}
\caption{\label{Fig15} The critical current density versus magnetic field curves of 5 K ${\rm <}$ T 
${\rm <}$ 35 K for the MgB$_{2}$ bulk sample  by the COSSET.}
\end{figure}

\newpage

\end{document}